\documentclass[twocolumn,showpacs,preprintnumbers,amsmath,amssymb]{revtex4}
\usepackage{graphicx}
\usepackage{dcolumn}
\usepackage{bm}
\begin{document}
\title{The electron-phonon theory of superconductors: vertex correction}
\author{Wei Fan}
\affiliation{Key Laboratory of Materials Physics, Institute of Solid
State Physics, Hefei Institutes of Physical Sciences, Chinese Academy
of Sciences, 230031-Hefei, People's Republic of China}
\date{\today}

\begin{abstract}
The strong coupling Eliashberg theory plus vertex correction is used
to calculate the maps of transition temperature (T$_{c}$) in
parameter-space characterizing superconductivity. Based on these
T$_{c}$ maps, complex crossovers are found when electron-phonon
interaction increases from weak-coupling region to strong coupling
region. Especially, the interplay interaction between vertex
correction and Coulomb interaction reduces the higher T$_{c}$
predicted in standard Eliashberg theory to the lower T$_{c}$ observed
in experiments. The doping-dependent T$_{c}$ of cuprate
superconductors can be explained as the effect of vertex correction in
standard Eliashberg theory and, most importantly, the pseudo-gap can
be explained as the effect of vertex correction. Our results are also
indicated that the non-adiabatic effect is the barrier for the
realizations of high T$_{c}$ in compounds with compositions of light
atoms and with high phonon frequencies.
\end{abstract}
\maketitle

\section{Introduction}

In the situation of very strong electron-phonon coupling,
non-adiabatic effects of electron-ion system will be so important that
the electrons are dressed heavily by lattice vibrations and the
conventional strong-coupling
theory\cite{Eliashberg1,Nambu1,Scalapino1,Allen1,McMillan1} need be
generalized to include the non-adiabatic effects or the vertex
corrects beyond Migdal's theorem. The strong-coupling theory including
vertex correction has been widely studied by using perturbation
theory~\cite{Kostur1,Grimaldi1,Mierzejewski1,Cappelluti1,Fan1} and
shown the existences of crossovers when electron-phonon interaction
evolving from weak-coupling region to strong coupling
region~\cite{Paci1,Capone1}. The problem of electron-phonon
interaction had been solved numerically using Quantum Monte Carlo
method (QMC)~\cite{Freericks1} and dynamic mean-field theory
(DMFT)~\cite{Freericks2,Hague1}. The crossover behavior was exhibited
in the calculation of Holstein-Hubbard model using QMC
method~\cite{Freericks1}. The crossover from electron behavior to
polaron behavior with increasing electron-phonon coupling had also
emerged from the path-integral calculations in theory of
polaron~\cite{Nasu1}, which had been used to explain the
superconductivity of copper-oxides high-temperature superconductors by
forming pairs of polarons~\cite{Alexandrov1,Alexandrov2,Nasu1}.

The standard strong-coupling theory has no bound on T$_{c}$. Recently,
the Eliashberg functions $\alpha^{2}F(\omega)$ extracted from the
measurements of optical spectroscopy for the iron based superconductor
Ba$_{0.55}$K$_{0.45}$Fe$_{2}$As$_{2}$~\cite{Yang1} and the
measurements of infrared optical conductivity for copper-oxides
superconductors~\cite{Heumen1} predicted very strong electron-phonon
interaction and very high T$_{c}$ over the experimental
values~\cite{Heumen1}. Additionally, the new found rich-hydrogen
superconductor silane (SiH$_{4}$)~\cite{Eremets1} is the realization
of the predicted high-temperature superconductor of metal hydrogen by
pre-compressed mechanism~\cite{Ashcroft1}. However, the T$_{c}$ of
silane is far lower than the predictions of standard string-coupling
theory~\cite{Chen_Wang}. What is the underlying reason of so high
T$_{c}$ in the predictions of the standard strong-coupling theory?.

In this paper, the T$_{c}$ maps including the influences of vertex
corrections are studied. Complex crossover are found on these T$_{c}$
maps when the parameter $\lambda$ of electron-phonon coupling
increases from weak-coupling region to strong-coupling region. These
crossovers are very close to the well known $\lambda$=2 at which the
value of T$_{c}$ reaches its maximum~\cite{McMillan1}. We also find
that T$_{c}$ does not monotonously increase with phonon frequency
$\Omega_{P}$. T$_{c}$ decreases with $\Omega_{P}$ when $\Omega_{P}$
higher than a threshold value. This means that high phonon-frequency
is unfavorable to superconductivity if vertex correction is strong. In
section~\ref{Discussion}, we will show that the vertex correction will
significantly suppress the predicted T$_{c}$ for
Ba$_{0.55}$K$_{0.45}$Fe$_{2}$As$_{2}$~\cite{Yang1}, the copper-oxides
superconductors~\cite{Heumen1} and superconducting
silane~\cite{Eremets1}.  In section~\ref{Result} main numerical
results will be presented. The basic theory used in this paper will be
introduced in section~\ref{Theory}.

\section{\label{Theory} Theoretical formulas}

The equation of energy gap in reference~\cite{Kostur1} had been
generalized by including the Coulomb interaction~\cite{Fan1}. The
electron-phonon interaction in Nambu's scheme is expressed as
\begin{equation}
H_{ep}=\sum_{kq\nu}\Psi^{+}_{k}\tau^{3}J_{k,q-k\nu}\Psi_{q-k}
(a^{+}_{-q\nu}+a_{q\nu}).
\end{equation}
\noindent where the index $\nu$ of phonon mode will be omitted if we
only consider a single phonon mode. We employ the isotropic
approximation and the coupling constant of electron-phonon interaction
is written as $J_{k,k'\nu}=J$. The calculations of vertex corrections
are greatly simplified under isotropic approximation because the
electron-phonon interactions are included in the vertex corrections
only by the functions of electron-phonon interaction $\lambda_{n}$
defined as
$\lambda_{n}=2\int^{\infty}_{0}d\omega\alpha^{2}F(\omega)\omega/(\omega^{2}+\omega_{n}^{2})$.
The energy-gap equation is expressed as
\begin{eqnarray}\label{GapEQ0}
Z_{n}&=&1+\frac{\pi}{|\omega_{n}|\beta}\sum_{n'}\lambda_{n-n'}A_{nn'}s_{n}s_{n'}a_{n'}
\\ \nonumber
Z_{n}\Delta_{n}&=&\frac{\pi}{\beta}\sum_{n'}[\lambda_{n-n'}B_{nn'}-\mu^{*}+C_{nn'}]
\frac{a_{n'}\Delta_{n'}}{|\omega_{n'}|}
\end{eqnarray}
\noindent with the renormalized function $Z_{n}\sim
Z_{k}(i\omega_{n})$, the energy-gap function
$\Delta_{n}\sim\phi_{k}(i\omega_{n})/Z_{k}(i\omega_{n})$ and the
parameters $A_{nn'}=1-V^{A}_{nn'}$, $B_{nn'}=1-V^{B}_{nn'}$,
$s_{n}=\omega_{n}/|\omega_{n}|$ and
$a_{n}=(2/\pi)\arctan(E_{B}/Z_{n}|\omega_{n}|)$. The three parameters
of vertex correction $V^{A}_{nn'}$, $V^{B}_{nn'}$ and $C_{nn'}$ can be
found in Ref.\cite{Fan1}. The Coulomb pseudo-potential is defined as
$\mu^{*}=\mu_{0}/(1+\mu_{0}\ln(E_{B}/\omega_{0}))$, where
$\mu_{0}=N(0)U$, U the Coulomb interaction between electrons and
$\omega_{0}$ characteristic energy of typical phonon correlated to
superconductivity.

If considering that $\Delta_{n}\rightarrow 0$ when temperature is very
close to T$_{c}$, the terms proportional to $\Delta_{n}^{2}$ are
ignored. The energy-gap equation is simplified to
$\sum_{n'=-\infty}^{+\infty}K_{nn'}(\Delta_{n'}/|\omega_{n'}|)=0$. The
kernel matrix is expressed as
\begin{eqnarray}\label{GapKN}
 K_{nn'}&=&[\lambda_{n-n'}B_{nn'}-\mu^{*}+C_{nn'}]a_{n'}-\delta_{nn'}H_{n'},
 \\ \nonumber
 H_{n'}&=&\sum_{n''=-\infty}^{+\infty}[
 \frac{\delta_{n'n''}|\omega_{n''}|}{\pi
 k_{B}T}+\lambda_{n'-n''}A_{n'n''}s_{n'}s_{n''}a_{n''}].
 \end{eqnarray}
\noindent If the vertex corrections are ignored, three parameters
$V^{A}_{nn'}$, $V^{B}_{nn'}$ and $C_{nn'}$ are all equal to zero and
the kernel Eq.(\ref{GapKN}) of energy-gap equation reduces to the
general form without vertex correction~\cite{Allen1} after some
symmetrizations and simplifications. In the calculation of $a_{n}$,
$Z_{n}\sim$1 is the value of normal state. It's convenient that the
$K_{nn'}$ matrix is symmetrized as in Ref.\cite{Allen1}. The T$_{c}$
is defined as the temperature when the maximum of eigenvalues
E$^{max}$ of kernel matrix $K_{nn'}$ crosses zero and changes its
sign. About $N$=200 Matsubara's energies are used to solve above
equation. Only 20-30 iterations are enough to search T$_{c}$ from -600
K to 600 K with accuracy 0.0001 K by using the bisection method.

In calculations of $\alpha^{2}F(\omega)$, the function
$\alpha^{2}(\omega)$ is approximately a constant around the peak of
phonon mode and the density of state of phonon is expressed as
\begin{eqnarray}\label{AFEQ}
 F(\omega)=\left\{
 \begin{tabular}{cc}
  $\frac{c}{(\omega-\Omega_{P})^{2}+(\omega_{2})^{2}}
  -\frac{c}{(\omega_{3})^{2}+(\omega_{2})^{2}}$, &
  $|\omega-\Omega_{P}|<\omega_{3}$ \\
  0 & others,
 \end{tabular}
 \right.
 \end{eqnarray}
\noindent where $\Omega_{P}$ is the energy of phonon mode,
$\omega_{2}$ the half-width of peak of phonon mode and
$\omega_{3}=2\omega_{2}$. The parameter of electron-phonon interaction
is defined as
\begin{equation}\label{lam}
\lambda=\lambda_{0}=2\int_{0}^{\infty}d\omega\alpha^{2}F(\omega)/\omega
.
\end{equation}
\noindent The well known value $\lambda=2$ given by McMillan measures
the instability of superconductivity induced by lattice instability
and plays very important role in this paper.

Experimental phonon spectrum and the phonon spectrum from
linear-response calculation~\cite{Savrasov1} are multi-peak structures
with broad energy distributions. To make the approximation of single
peak model to be more reliable approximation, the effective phonon
frequency or energy should be used. A good choice is the
$\langle\omega\rangle_{ln}=\exp(2/\lambda\int
d\omega\ln(\omega)\alpha^{2}F(\omega)/\omega)$ defined in
Ref.~\cite{Allen1}. The $\langle\omega\rangle_{ln}$ is calculated from
Eliashberg function that can be obtained from linear-response theory
or extracted from the experimental measurements of phonon properties.

\section{\label{small} Small parameter for perturbing calculation}

It's important to known the small parameter for perturbing
calculations in the theory of electron-phonon interaction. The matrix
element of electron-phonon interaction is given by
$J_{k,k'\nu}=(\hbar/2M\omega_{k-k'})^{1/2}\Pi_{\nu}(k,k')$
($\hbar=1$). The forms of $\Pi_{\nu}(k,k')$ are dependent on the kinds
of electron-phonon: deformation potential or polar
coupling~\cite{Mahan1}. Under isotropic approximation, the expression
of electron-phonon coupling constant is given by
\begin{equation}\label{EPCJ}
J=\sqrt{\frac{\lambda}{2}\frac{\Omega_{P}}{N(0)}}
\end{equation}
if the well-known relation
$M\lambda\langle\omega^{2}\rangle=N(0)\langle \Pi^{2}\rangle_{FM}$ and
the Einstein spectrum of phonon are used, where $\Omega_{P}$ is phonon
energy, $N(0)$ is density of state at Fermi energy,
$\langle\rangle_{FM}$ is the averages of Fermi surface and
$\langle\rangle$ is the average weighted by
$(2/\lambda)\alpha^{2}F(\omega)/\omega$. The Eliashberg function with
Einstein spectrum is expressed as
$\alpha^{2}F(\omega)=(\lambda/2)\omega\delta(\omega-\Omega_{P})$. The
small parameter for perturbing calculation is
\begin{equation}\label{small_SC}
J/E_{B}=\sqrt{\frac{\lambda}{2h}\frac{\Omega_{P}}{E_{B}}}.
\end{equation}
\noindent where dimensionless parameter $h=N(0)E_{B}$. The standard
strong-coupling theory is correct only $\Omega_{P}/E_{B}\ll 1$ (
Migdal's theorem ). We can see that $\lambda\ge 2$ is equivalent to
$J^{2}>\Omega_{P}E_{B}/h$. The small parameters $J/E_{B}$$<$0.3 and
$\Omega_{P}/E_{B}\simeq 0.30-0.50$ for the fullerides such as
Rb$_{3}$C$_{60}$ are estimated by using Eq.(\ref{small_SC}) with the
parameters $N(0)$$\sim$8 (states/eV), $E_{B}$$\sim$0.5 eV,
$\lambda$$\sim$0.5-1.0 and $\Omega_{P}$$\sim$65-100 meV. Thus
strong-coupling theory plus vertex correct is suitable for fullerides.

As a comparison, the effective interaction of electrons in BCS
theory~\cite{Bardeen1} can be written as
\begin{equation}\label{BCSV1}
\frac{1}{2}\sum_{kq}(v_{q}+
\frac{2\omega_{q}|J_{k,k+q}|^{2}}{(E_{k+q}-E_{k})^{2}-\omega_{q}^{2}})
c^{\dagger}_{k+q,\sigma}c^{\dagger}_{-k-q,-\sigma}c_{-k,-\sigma}c_{k,\sigma}.
\end{equation}
\noindent Under isotropic approximation, by using
$\omega_{q}\sim\Omega_{P}$, $E_{k+q}\sim E_{k}\sim E_{F}$, $v_{q}=U$
and Eq.(\ref{EPCJ}), the above equation can be simplified as
\begin{equation}\label{BCSV2}
-V_{BCS}\sum_{kq}
c^{\dagger}_{k+q,\sigma}c^{\dagger}_{-k-q,-\sigma}c_{-k,-\sigma}c_{k,\sigma}
\end{equation}
\noindent where $V_{BCS}=(\lambda-\mu^{*})/2N(0)$. The small parameter
for BCS theory is
\begin{equation}\label{small_BCS}
V_{BCS}/E_{B}=\frac{\lambda-\mu^{*}}{2h}.
\end{equation}
\noindent The perturbation theory is correct only
$(\lambda-\mu^{*})/2h<1$ or $\lambda<2h+\mu^{*}$, however
$\lambda>\mu^{*}$ to keep superconducting state stable. If
$\lambda>2h+\mu^{*}$, the strongly coupling electron-pairs will
significantly modify the electronic structure of superconductors and
lead to structural instability.

\section{\label{Result} The general results of vertex corrections}

The three-dimensional T$_{c}$ maps had been calculated in the previous
paper~\cite{Fan1}. In this section, three-dimensional T$_{c}$ maps in
$\lambda$-$\Omega_{P}$-$\mu^{*}$ phase space including vertex
corrections are calculated by using the simple phonon spectrum with
the form of Eq.(\ref{AFEQ}). From electron point of view, the vertex
correction or non-adiabatic effect can be controlled by the effective
width E$_{B}$, on the other hand, from ion point of view, it can be
controlled by the cutoff of phonon energy $\omega_{0}$. In this work,
the vertex corrections are controlled by the effective band-width
E$_{B}$ with the range from 0.5 eV to 5 eV. The situation
E$_{B}=\infty$ is equivalent to no vertex correction.

\begin{figure}
\includegraphics[width=0.40\textwidth]{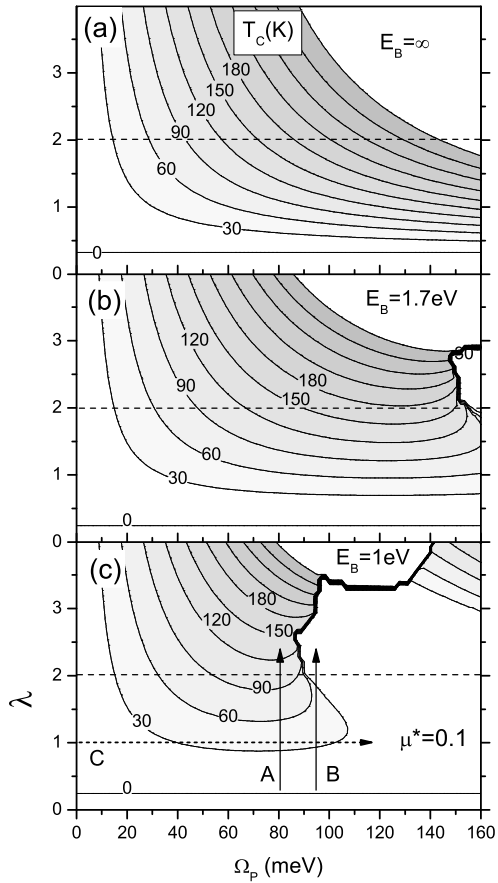}
\caption{\label{fig1}The evolution of T$_{c}$ map on
$\lambda$-$\Omega_{P}$ plane with increasing strengths of vertex
corrections (decreasing effective band-width E$_{B}$) with (a)
E$_{B}=\infty$, (b) E$_{B}$=1.7 eV and (c) E$_{B}$=1 eV. The Coulomb
pseudo-potential $\mu^{*}$=0.1. }
\end{figure}

\begin{figure}
\includegraphics[width=0.40\textwidth]{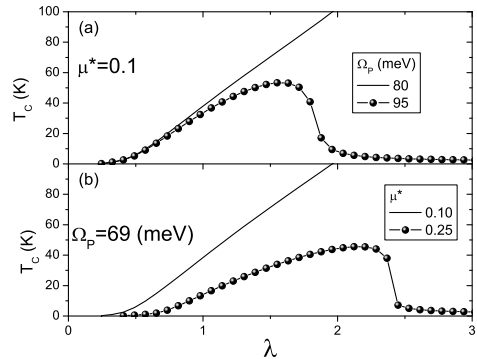}
\caption{\label{fig2}(a) The T$_{c}$ change along two arrows shown in
Fig.\ref{fig1}(c) with fixed phonon energies $\Omega_{P}$=80 meV and
95 meV respectively. (b) The T$_{c}$ change along two arrows shown in
Fig.\ref{fig3}(c) with fixed Coulomb pseudo-potentials $\mu^{*}$=0.10
and 0.25 respectively.}
\end{figure}

\begin{figure}
\includegraphics[width=0.40\textwidth]{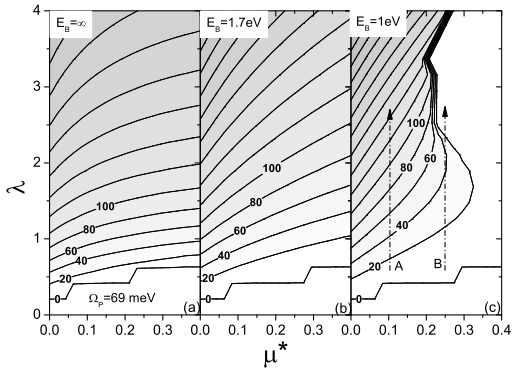}
\caption{\label{fig3}The evolutions of T$_{c}$ map on
$\mu^{*}$-$\lambda$ plane ($\Omega_{P}$=69 meV) with decreasing
effective band-width (a) E$_{B}=\infty$, (b) E$_{B}$=1.7 eV and (c)
E$_{B}$=1.0 eV.}
\end{figure}

The Fig.\ref{fig1}(a,b,c) illustrate the evolution of T$_{c}$ map with
decreasing E$_{B}$. The Fig.\ref{fig1}(a) is the T$_{c}$ map having
been obtained in the previous work without considering vertex
corrections~\cite{Fan1}. The large deformation with strong vertex
correction starts at E$_{B}$=1.7 eV in Fig.\ref{fig1}(b) near the well
known $\lambda$=2.0 and high phonon energy. With E$_{B}$ decreasing to
1 eV further, the region with strong vertex correction rapidly expands
and occupies large part of parameter space with $\Omega_{P}$$>$80 meV
in Fig.\ref{fig1}(c). In the region with $\Omega_{P}$$<$80 meV, the
T$_{c}$ is strongly suppressed however there are no discontinuous
changes of T$_{c}$ or breaking of contour lines. An important result
from the Fig.\ref{fig1} is that T$_{c}$ does not change with $\lambda$
monotonously if phonon energy $\Omega_{P}$ is high enough. The
Fig.\ref{fig2}(a) shows the changes of T$_{c}$ with $\lambda$ along
two arrows (A,B) shown in Fig.\ref{fig1}(c). If $\Omega_{P}$=80 meV,
the T$_{c}$ monotonously increases with $\lambda$. However for
$\Omega_{P}$=90 meV, the T$_{c}$ first increases with $\lambda$,
reaches the maximum at $\lambda\sim 1.5-1.7$ and then quickly
decreases with increasing $\lambda$. Further increasing $\lambda$$>$2,
T$_{c}$ will be very low due to strong vertex corrections. The
non-monotonous $\lambda$-dependent T$_{c}$ in Fig.\ref{fig2}(a) had
been found in the non-adiabatic theory of
superconductivity~\cite{Paci1}. Some crossover behaviors from weak
coupling to strong coupling region had been predicted in
Holstein-Hubbard model solved numerically by quantum Monte Carlo
method~\cite{Freericks1} and in polaron theory~\cite{Nasu1}. It's very
reasonable that the non-monotonous $\lambda$-dependent T$_{c}$ is
equivalent to the crossovers found in QMC
calculation~\cite{Freericks1} and polaron theory~\cite{Nasu1}. So only
the leading vertex correction can describe qualitatively very well the
electron-phonon interaction in strong coupling region.

In the previous paper, the Coulomb pseudo-potential $\mu^{*}$ had
small effects when $\mu^{*}$$>$0.2 in the calculations without vertex
corrections~\cite{Fan1}. The Fig.\ref{fig3}(a) is the normal T$_{c}$
map on $\mu^{*}$-$\lambda$ plane without vertex
correction~\cite{Fan1}. The figure shows that when $\mu^{*}$$>$0.2,
T$_{c}$ is insensitive to the change of $\mu^{*}$. The breaking
contour line with T$_{c}$=0 K are because of the inaccurate
calculations when T$_{c}$$<$0.1 K if only $N$=200 Matsubara energies
are used. The contour lines with T$_{c}$$>$0.1 K is accurate enough.
If the Coulomb pseudo-potential and vertex correction work together,
the situation will change drastically and some new interesting results
will appear. The large deformations are found in Fig.\ref{fig3}(c) if
E$_{B}$ decreases to 1.0 eV. In the region with $\lambda$$<$2.0, the
contour lines with higher T$_{c}$ are moved out but the contour lines
with lower T$_{c}$ fill in this region. As expected, the large
deformations and discontinuous changes of contour lines appear on the
T$_{c}$ map when $\mu^{*}$$>$0.20. The contour lines with iso-values
from T$_{c}$=20 K to 200 K are packed together within the region
0.20$<$$\mu^{*}$$<$0.25 and $\lambda$$>$2. The figure clearly shows
that if the Coulomb pseudo-potential $\mu^{*}$ is larger enough, the
T$_{c}$ will change with $\lambda$ non-monotonously.  The changes of
T$_{c}$ along two arrows with $\mu^{*}$=0.1 and 0.25 are plotted in
Fig.\ref{fig2}(b). For $\mu^{*}$=0.25, T$_{c}$ first increases with
$\lambda$ until reaches the maximum at $\lambda$=2.2 and then sharply
decreases to smaller value at $\lambda$=2.5. The crossover behavior is
enhanced by strong Coulomb interaction.

\begin{figure}
\includegraphics[width=0.40\textwidth]{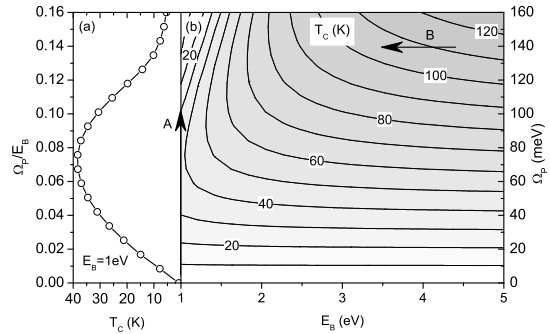}
\caption{\label{fig4} (b) T$_{c}$ map on $E_{B}$-$\Omega_{P}$ plane
with $\mu^{*}$=0.1 and $\lambda$=1.0. (a) T$_{c}$ changes
non-monotonously with increasing $\Omega_{P}/E_{B}$ along direction of
arrow $A$ with increasing $\Omega_{P}$ and fixed E$_{B}$=1.0 eV.}
\end{figure}

There are three methods that can control vertex correction: (1) only
$\omega_{0}$ changes and E$_{B}$ keeps unchanged, (2) $E_{B}$ changes
and $\omega_{0}$ keeps unchanged , (3) both $E_{B}$ and $\omega_{0}$
change. It's possible that three methods provide qualitatively
different results. In this paper, we use narrow peak approximation so
that $\omega_{0}\sim\Omega_{P}$. The T$_{c}$ map on
$\Omega_{P}$-$E_{B}$ plane is plotted in Fig.\ref{fig4}(b). The
Fig.\ref{fig4}(a) shows how T$_{c}$ changes with $\Omega_{P}/E_{B}$
along the direction of arrow $A$ in Fig.\ref{fig4}(b) with E$_{B}$=1
eV. It's found that at first T$_{c}$ increases with $\Omega_{P}/E_{B}$
and then decreases with it. The arrow $A$ in Fig.\ref{fig4}(b) is
corresponding to the horizontal arrow $C$ in Fig.\ref{fig1}(c). The
behavior of T$_{c}$ with $\omega_{0}/E_{F}$ had been found in a series
of references~\cite{Grimaldi1,Cappelluti1,Paci1} in that coupling
constant of electron-phonon interaction $g_{k,k+q}$ ( or $J_{k,k+q}$
in this paper) is dependent on a cutoff $q_{c}$ for wave-vector $q$.
From Fig.\ref{fig1}(c) and Fig.\ref{fig4}(b), the increasing T$_{c}$
at small $\Omega_{P}/E_{B}$ isn't the effect of vertex correction and
is the standard result of strong-coupling theory. The real effects of
vertex corrections are that T$_{c}$ decreases with $\Omega_{P}/E_{B}$
at large $\Omega_{P}/E_{B}$. Certainly, the vertex correction is
enhanced if $\Omega_{P}/E_{B}$ increases by decreasing E$_{B}$ with
fixed $\Omega_{B}$, illustrated by the horizontal arrow $B$ in
Fig.\ref{fig4}(b).

\begin{figure}
\includegraphics[width=0.40\textwidth]{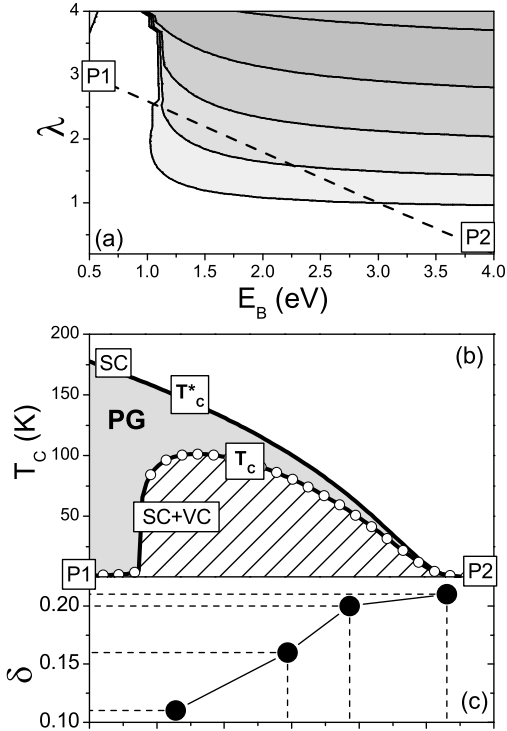}
\caption{\label{fig5}(a) The T$_{c}$ map on $E_{B}$-$\lambda$ plane
with $\mu^{*}$=0.25 and $\Omega_{P}$=72 meV. (b) The open circle line
is the evolution of T$_{c}$ from $P1$ to $P2$ in (a). The solid line
T$_{c}^{*}$ is the standard results in strong coupling theory without
vertex correction. (c) The $\delta$-$\lambda$ relation is adopted in
Ref.\cite{Heumen1}}
\end{figure}

Finally, T$_{c}$ map on E$_{B}$-$\lambda$ plane is presented in
Fig.\ref{fig5} with $\Omega_{P}$=72 meV. If E$_{B}$ increases but
$\lambda$ keeps unchanged, the T$_{c}$ monotonously increases with
E$_{B}$ until to the limit of non-vertex correction. More
interestingly, on this map, the T$_{c}$ is non-monotonous dependent on
E$_{B}$ along strait line from $P1$ to $P2$ companying by the decrease
of $\lambda$ from 3.0 to 0.2. The non-monotonous dependence of T$_{c}$
on effective band-width E$_{B}$ is similar to the band-filling effects
of T$_{c}$ in the non-adiabatic theory of
superconductivity~\cite{Cappelluti1}. Our results show that, if
$\Omega_{P}>$80 meV, the suppression of T$_{C}$ will be more
prominently.

\section{\label{Discussion} Discussion}

As mentioned in the introduction, the values of T$_{c}$ obtained from
standard strong-coupling theory are generally higher than those
measured in experiments. The copper-oxides superconductors
Bi$_{2}$Sr$_{2}$CaCu$_{2}$O$_{8+\delta}$ and
Bi$_{2}$Sr$_{2}$Cu$_{2}$O$_{6+\delta}$ studied in Ref~\cite{Heumen1}
have very strong electron-phonon interactions $\lambda$$\sim$2.36-2.85
and overestimated T$_{c}$ in underdoped samples. With increasing
doping $\delta$, the values of $\lambda$ decrease to about 0.35-1.42.
The effective band-widths E$_{B}$ of conducting electrons for these
cuprates are distributed from 1 eV to 3 eV. The effective phonon
energies are distributed from 50 meV to 80 meV. The Coulomb
interactions are strong in cuprate superconductors $\mu^{*}$=0.25. We
can clearly see from Fig.\ref{fig2}(b) and Fig.\ref{fig3}(c) that in
underdoped samples, T$_{c}$ can be reduced to very small values in
strong electron-phonon interaction region ($\lambda$$>$2.5) due to the
interplay interaction between vertex correction and strong Coulomb
interaction. As shown in Fig.\ref{fig5}(b), the values of T$_{c}$ for
all samples are reduced from around 200 K to lower than 150 K and
close to experimental values~\cite{Heumen1}. Our results are provided
an explanation to pseudo-gap in underdoped region shown in
Fig.\ref{fig5}(b). The cooper-pairs pre-form at the mean field value
T$_{c}^{*}$ of transition temperature in standard strong-coupling
theory. However strong non-adiabatic effects induce the instability of
cooper-pairs and suppress real T$_{c}$ to lower values. The
T$^{*}_{c}$ degenerates with T$_{c}$ in overdoped region is similar to
the example (1) of Fig.9 in Ref.\cite{Norman1}.

For the iron-based superconductors
Ba$_{0.55}$K$_{0.45}$Fe$_{2}$As$_{2}$ studied in Ref.\cite{Yang1}, the
parameters of boson-phonon interaction decrease from 3.42 to 0.78 with
increasing temperature. The contributions of electron-phonon
interactions have energies lower than 40 meV in Eliashberg functions
$\alpha^{2}F(\omega)$. At 28 K close to T$_{c}$, the roughly estimated
value of $\lambda$ contributed by energies lower than 40 meV is about
2.0 close to the values at temperature 86 K. This is too strong to
account to experimental T$_{c}$=28 K. Because the effective phonon
energy is about 20-40 meV, the reduced T$_{c}$ from the vertex
corrections are also small (Fig.\ref{fig1}) although the effective
band-widths of conducting electrons are small from 1 eV to 2 eV. The
key problem is that how we can extract the more accurate value of
$\lambda$ of electron-phonon interaction from total $\lambda$. If we
use the $\lambda$=1.44 obtained at temperature 151 K and the effective
phonon energy $\Omega_{P}$=20 meV, the obtained T$_{c}$=29 K is close
to experimental T$_{c}$=28 K. The reasonable value for iron-based
superconductor is about $\lambda$=1.0. This is indicative that the
contribution of electron-phonon interaction to Eliashberg function
$\alpha^{2}F(\omega)$ dominates over other interactions at high
temperature.

From Fig.\ref{fig4}, we can see that if effective band-width E$_{B}$
is small, the T$_{c}$ will decrease with increasing phonon frequency
if $\Omega_{P}$ is higher enough. This means that we can not increase
T$_{c}$ infinitely by decreasing mass of atom or increasing phonon
frequency. The strong vertex corrections in high phonon-energy region
suppress T$_{c}$ to smaller values than those of predicted by standard
strong-coupling theory. This is a reasonable explanation for the lower
T$_{c}$ in the experiment of silane superconductor~\cite{Eremets1}
than that in the strong-coupling theory~\cite{Chen_Wang}.

The superconductivity of fullerides can be described using
Cooper-pairs glued by virtual phonon excitation. However there are
some very important problems are unsolved~\cite{Gunnarsson1}. The
fullerides have very large number of phonon modes from low frequency
to high frequency. What phonon modes are most important to their
superconductivity?. There are two groups of phonon modes of
fullerides: (a) inter-molecular molecular modes with energies lower
than 150cm$^{-1}$ (or 18 meV) and (b) intra-molecular phonon modes
with energies higher than 250 cm$^{-1}$ (or 31 meV). The
inter-molecular modes have small contributions to T$_{c}$ due to their
lower energies and not too strong electron-phonon interaction. Most
importantly, our results shown in Fig.\ref{fig1}(c) and Fig.\ref{fig4}
prove that the `tangential' intra-molecular modes with energy higher
than 1000 cm$^{-1}$ (or 124 meV) have small contributions to T$_{c}$
because of strong vertex corrections. So the intra-molecular `radial'
modes from 250cm$^{-1}$ to 1000cm$^{-1}$ of group (b) should
contribute main parts of superconductivity. In fact effective phonon
frequency $\langle\omega\rangle_{ln}$ is located in this energy range.
The empty t$_{1u}$ orbits will form narrow energy bands with width
about 500 meV in C$_{60}$ solid and strongly couple with
intra-molecular H$_{g}$ phonons. In Rb$_{3}$C$_{60}$ solid, the width
of t$_{1u}$ bands increase to about 1 eV and is half filled with
electrons. Based on our calculations using density functional theory
with plane-wave pseudo-potential methods, the effective band-width
E$_{B}$ is higher than 500 meV and lower than 1 eV. If we choose the
intra-molecular radial mode $\Omega_{P}$=525 cm$^{-1}$ (or 65 meV) and
the half-width E$_{B}$=500 meV, T$_{c}$ will be lower than
experimental value due to vertex-correction effects shown in
Fig.\ref{fig5}(a). If we slightly increase E$_{B}$ and use the
averaged value E$_{B}$=750 meV, both T$_{c}$=29.5 K and isotope
coefficient $\alpha\sim0.30$~\cite{Chen1} are close to the values in
experiments if $\lambda$=1.04 and $\mu^{*}$=0.1. The required
parameter $\lambda$ is different from our previous work value
0.66~\cite{Fan2} in the calculation without vertex correction, however
still close to the reasonable range from 0.5 to
1.0~\cite{Gunnarsson1}. The accurate calculations of $\lambda$ and
$\mu^{*}$ are important to understand the role of vertex correction.

\section{Conclusion}

In summary, the strong-coupling Eliashberg theory including vertex
correction is systemically studied in this paper. The T$_{c}$ maps in
parameter-space $\lambda$-$\Omega$-$\mu^{*}$ for different E$_{B}$
contain the completed information on strong-coupling theory and the
vertex corrections.  Especially, the combined interaction of vertex
correction and Coulomb interaction can significantly depress T$_{c}$
to small value. The non-monotonous changes of T$_{c}$ with increasing
$\lambda$ show the crossover behaviors near $\lambda$=2 when $\lambda$
evolving from weak-coupling region to strong-coupling region. The
crossovers can explain the doping-dependent T$_{c}$ of cuprate
superconductors if the Coulomb interactions are strong. The crossover
behavior in $\Omega_{P}$-dependent T$_{c}$ indicates that
high-frequency phonon is unfavorable to high T$_{c}$. Thus the strong
non-adiabatic effect makes it hard to find high T$_{c}$
superconductors in compounds containing light elements. Finally the
T$_{c}$ maps in the previous paper~\cite{Fan1} and the maps with
vertex corrections in this paper provide very comprehensive
understanding of superconductivity of superconductors.

\section{Acknowledgement}

The author thanks Prof. E. Cappelluti for very helpful discussions.
This work is supported by Director Grants of Hefei Institutes of
Physical Sciences, Knowledge Innovation Program of Chinese Academy of
Sciences and National Science Foundation of China.

\end{document}